\documentclass[sigconf, screen, nonacm]{acmart}

\AtBeginDocument{%
  }

\usepackage{amsmath}

\DeclareMathOperator*{\argmin}{argmin}
\usepackage{booktabs}
\usepackage{subfigure}
\usepackage{eurosym}
\usepackage{xurl}     % Better URL breaking (works with hyperref)
\hypersetup{breaklinks=true} % Enables line breaks in links

\newcommand{\R}{\ensuremath{\mathbb{R}}}
\usepackage[nonumberlist, acronym]{glossaries}
\makeglossaries
\newacronym{hvac}{HVAC}{Heating, Ventilation, and Air-Conditioning}
\newacronym{mae}{MAE}{Mean Absolute Error}
\newacronym{mse}{MSE}{Mean Squared Error}
\newacronym{dfl}{DFL}{Decision-Focused Learning}
\newacronym{ito}{ITO}{Identify-Then-Optimize}
\newacronym{rl}{RL}{Reinforcement Learning}
\newacronym{rc}{RC}{Resistance-Capacitance}
\newacronym{spo}{SPO}{Smart "Predict, then Optimize"}
\newacronym{spo+}{SPO+}{Smart "Predict, then Optimize" Plus}
\newacronym{sgd}{SGD}{Stochastic Gradient Descent}
\newacronym{ahu}{AHU}{Air Handling Unit}
\newacronym{ml}{ML}{Machine Learning}
\newacronym{lp}{LP}{Linear Programs}
\newacronym{cp}{CP}{Combinatorial Programs}
\newacronym{kkt}{KKT}{Karush-Kuhn-Tucker}
\newacronym{snr}{SNR}{Signal-to-Noise Ratio}

\begin{document}

\title[DFL for Complex System Identification: HVAC Management System Application]{Decision-Focused Learning for Complex System Identification: HVAC Management System Application}

\author{Pietro Favaro}
\orcid{0009-0002-0782-4494}
\author{Jean-François Toubeau}
\orcid{0000-0001-9853-2694}
\author{François Vallée}
\orcid{0000-0002-2409-2128}

\affiliation{%
  \department{Power Systems and Markets Research Group}
  \institution{University of Mons}
  \city{Mons}
  \country{Belgium}
}
\email{pietro.favaro@umons.ac.be}
\email{jean-francois.toubeau@umons.ac.be}
\email{francois.vallee@umons.ac.be}

\author{Yury Dvorkin}
\affiliation{%
  \department{Ralph O’Connor Sustainable Energy Institute}
  \department{Department of Electrical and Computer Engineering}
  \department{Department of Civil and System Engineering}
  \institution{Johns Hopkins University}
  \city{Baltimore}
  \state{Maryland}
  \country{US}}
\email{ydvorki1@jhu.edu}

\renewcommand{\shortauthors}{Favaro et al.}

\begin{abstract}
As opposed to conventional training methods tailored to minimize a given statistical metric or task-agnostic loss (e.g., mean squared error), Decision-Focused Learning (DFL) trains machine learning models for optimal performance in downstream decision-making tools.
We argue that \acrshort{dfl} can be leveraged to learn the parameters of system dynamics, expressed as constraint of the convex optimization control policy, while the system control signal is being optimized, thus creating an end-to-end learning framework. This is particularly relevant for systems in which behavior changes once the control policy is applied, hence rendering historical data less applicable. The proposed approach can perform system identification --- \textrm{i.e.}, determine appropriate parameters for the system analytical model --- and control simultaneously to ensure that the model's accuracy is focused on areas most relevant to control.
Furthermore, because black-box systems are non-differentiable, we design a loss function that requires solely to measure the system response. We propose pre-training on historical data and constraint relaxation to stabilize the \acrshort{dfl} and deal with potential infeasibilities in learning.
We demonstrate the usefulness of the method on a building Heating, Ventilation, and Air Conditioning day-ahead management system for a realistic 15-zone building located in Denver, US. The results show that the conventional \acrshort{rc} building  model, with the parameters obtained from historical data using supervised learning, underestimates HVAC electrical power consumption. For our case study, the ex-post cost is on average six times higher than the expected one. Meanwhile, the same RC model with parameters obtained via \acrshort{dfl} underestimates the ex-post cost only by 3\%.
\end{abstract}




\maketitle

\setlength{\glsdescwidth}{0.8\textwidth}
\printglossary[type=\acronymtype, title=List of Acronyms, nogroupskip=true, style=super] %, style=long/tree/long4col

\section{Introduction}
\label{sec: introduction}

Optimization is used to devise control policies such as for Heating, Ventilation, and Air Conditioning (HVAC) in buildings \cite{atam_convex_2015}, the guidance of vehicles \cite{wang_survey_2024}, and the planning of complex dynamic systems such as day-ahead energy scheduling of active buildings \cite{s_m_hosseini_robust_2019}, or energy management systems for micro-grid users \cite{phani_raghav_optimal_2022}. The optimization problem solution is the actions or decisions to apply to the system. However, an analytical model of the system dynamics is always assumed to be known and formulated within an optimization problem, usually as constraints. The necessity to have an analytical model hinders control tasks where physics-based models are unpractical; either because the system is a black box, or physical models are far too complex. Data-driven surrogates become impractical when system dynamics change drastically due to a newly applied control policy, rendering prior historical observations incomplete and minimally informative. Therefore, the conventional two-stage Identify-Then-Optimize (ITO) approach is inefficient.
\par
This paper addresses the \acrshort{ito} approach inefficiencies using the day-ahead \acrshort{hvac} scheduling in buildings as a use case. \acrshort{hvac} scheduling becomes increasingly important because of the roll-out of smart meters and dynamic electricity tariffs \cite{zhou_comprehensive_2023}.
% Electricity prices are governed by the day-ahead market and vary hourly or quarterly \cite{shah_comprehensive_2020}. 
The underlying goal of such dynamic (e.g., time-of-use) tariffs is to prompt more flexibility at the consumption level to accommodate more renewable-based, and thus often weather-dependent and uncertain, generation. Unlike lightning or essential equipment that needs to be available on demand, \acrshort{hvac} can be scheduled and benefit from the building thermal inertia to shift the consumption across the day. Consequently, building managers can save money by optimizing \acrshort{hvac} operations. Moreover, the building sector accounts for 75\% of the final electricity consumption in the USA, in which \acrshort{hvac} consumption contributes  40\% \cite{shoemaker_nrel_2023, gonzalez-torres_review_2022}. Therefore, if minimizing greenhouse gas emissions, \acrshort{hvac} scheduling can have a significant impact.

\subsection{Building Thermodynamics Modeling}
The goal of day-ahead \acrshort{hvac} scheduling is to find the optimal temperature set point profiles for each building thermal zone to minimize the electricity cost for the upcoming day, alleviate the potential strain on the power grid, and ensure  thermal comfort. Therefore, building management systems rely on a thermodynamic model of the building that relates the \acrshort{hvac} operation to the indoor temperature.
\par
Historically, the thermodynamic model has always been assumed to be known. Most \acrshort{hvac} models are physics-based (\textrm{i.e.}, white-box) \cite{afroz_modeling_2018}. These models contain major assumptions. Moreover, some building parameters, such as the materials resistance and capacitance, are intrinsically unknown because of aging and installation process which may cause large error \cite{dagostino_experimental_2022}. Last, it requires expertise in \acrshort{hvac} modeling, which may not be readily available at each building site. \newline
On the other end of the spectrum lies fully data-driven models (\textrm{i.e.}, black-box). Neural networks were widely used for system identification \cite{drgona_physics-constrained_2021}, but cannot be easily integrated into optimization models; otherwise, the problem becomes non-linear and there is no guarantee on the solution quality \cite{huang_model_2014, zeng_predictive_2015}. Furthermore, data-driven models perform poorly out of their training zone \cite{afroz_modeling_2018}. This hinders day-ahead \acrshort{hvac} scheduling applications since their goal is to explore the whole feasible domain to find the best course of action. Reinforcement Learning (RL) has proven to be effective for real-time set point control \cite{yu_review_2021}, emission minimization \cite{jeen_low_2023}, and peak demand reduction \cite{sun_continuous_2020}. However, inputs and assumptions underlying \acrshort{rl} methods are often impractical. For instance, the active \acrshort{rl} algorithm in \cite{Jang2023ActiveRL} assumes real-time measurement of the occupants' clothing value, number, and thermal comfort. A review concludes that model-based deep RL must be favored to leverage some prior knowledge \cite{yu_review_2021}. Similarly, the review by \cite{homod_review_2013} states that the grey-box models are the most promising.\newline
Grey-box models involve physics-based formulation in which parameters are obtained through data-driven techniques. Grey-box models require less data than black-box models \cite{afram_review_2014}. Among such models, the Resistance-Capacitance (RC) model (also named lumped capacitance model or network model), a reduced order physics model, has exhibited good performances for buildings where there is no important indoor air convection \cite{kircher_lumped_2015}. Furthermore, the \acrshort{rc} model is linear, thus lending convexity to the optimization problem. However, estimating the parameters of the model is challenging. A first option to estimate the parameters is to analyze the building materials and select default tabular data. In addition to requiring building blueprints, this method does not account for manufacturing dispersion (\textrm{i.e.}, variations in product dimensions, properties, or performance due to inconsistencies in the production process), the on-site installation process, and the aging of materials. Therefore, data-driven parameter identification has led to significantly improved \acrshort{rc} models \cite{belic_thermal_2016}.

\subsection{Decision-Focused Learning}
In this paper, we adopt a classical \acrshort{rc} model. Aware of the limitations in estimating the parameters on historical data, we learn the parameters in a task-aware manner. In \cite{kao_directed_2009}, the authors were the first to suggest training a Machine Learning (ML) model in a directed manner based on its impact on the downstream decisions. The weights of the ML model that is used to predict the uncertain parameters of the downstream optimization problem are learnt to minimize the decision error induced by the parameters misestimation. To that end, the authors established the gradient of the optimization problem solution with respect to the problem parameters. This enables backpropagation through the optimization problem. Only unconstrained quadratic problems were considered. By allowing loss functions that depend explicitly on downstream optimization decisions (and implicitly on the ML model output), the calculation of optimization problem gradient paves the way for Decision-Focused Learning (DFL). The framework was then extended to stochastic quadratic problems \cite{donti_task-based_2017}. Because the gradient of unconstrained problems is easier to compute, the constraints were relaxed in the objective function.
\par
Bounded Linear Programs (LP) and Combinatorial Program (CP) lead to zero-valued gradient almost everywhere, and undefined elsewhere. Geometrically, the solution of \acrshort{lp} belongs to the set of the vertices of the feasible space polytope \cite{bazaraa_linear_2009}. Consequently, an infinitesimal change in the parameter values has no impact on solution; \textrm{i.e.}, the solution remain on the same vertex. However, as soon as parameter changes become large enough, the \acrshort{lp} solution jumps to another vertex. This results in a piecewise constant objective function, leading to a gradient that is either zero or undefined, making it uninformative for gradient descent training. For CP, the discrete nature of the decision variables leads to the same observation as for \acrshort{lp}. The first method to calculate the gradient of an \acrshort{lp} is to calculate the gradient of its Karush-Kuhn-Tucker (KKT) conditions. The gradient of the KKT conditions is also null or undefined since the KKT conditions are an exact representation of the same \acrshort{lp}. To address this issue, the objective function can be augmented by a L$_{\text{2}}$ regularization of the decisions variables. This turns the \acrshort{lp} into a strong concave or convex quadratic program if it is a maximization or minimization, respectively \cite{wilder_melding_2019}. The second approach is to design a surrogate loss function. Smart "Predict, then Optimize" Plus (SPO+) is such a convex function \cite{adam_n_elmachtoub_smart_2017}. Interestingly, SPO+ extend beyond \acrshort{lp} and can be applied to any optimization problem with a linear objective function and linear, convex, and integer constraints.
However, the uncertain parameters cannot appear in the constraints. Moreover, SPO+ requires the optimal decisions to be known, which may not always be the case. The third option for \acrshort{lp} and CP is stochastic smoothing. A random perturbation is applied to uncertain parameters to smooth the loss function transition and, thus, create informative gradients from the otherwise null and undefined derivative \cite{silvestri_score_2024}. Unlike SPO+, the uncertain parameters can appear in constraints.
\par
Leveraging the differentiability of conic programs, \citeauthor{agrawal_learning_2020} demonstrated how controller parameters within the objective function can be learned assuming full knowledge and differentiability of the system, and observation of the system dynamics \cite{agrawal_learning_2020}. It was applied to the forecasting of electricity prices through wind power prediction \cite{wahdany_more_2023}. It was also employed to split the demand for flexibility from the transmission grid towards the flexibility of the distribution grid assets \cite{ortmann_tuning_2024}.
\par
To the best of the authors' knowledge, no prior work has simultaneously addressed system identification (\textrm{i.e.}, learning constraint parameters, specifically the parameters of the dynamics model) and optimization, nor has it considered scenarios where the controlled system is both unknown and non-differentiable. Tackling this challenge requires a profound understanding of optimization and decision-focused learning theory, combined with extensive domain-specific expertise. The only previous work that does not assume knowledge of the system dynamics and differentiability is \cite{shah_decision-focused_2022} in which convex surrogate for both the optimization problem and the performance loss is constructed to train a \acrshort{ml} forecaster. However, the construction of the surrogate model requires knowing the true value of the uncertain parameters.

\subsection{Contributions}
This paper presents three major contributions.
\par
First, we leverage advances in decision-focused learning to perform complex system identification and control simultaneously. Compared to \cite{agrawal_learning_2020} and \cite{adam_n_elmachtoub_smart_2017}, uncertain parameters to be learned are within the constraints of the optimization control policy. Due to end-to-end learning, the system model parameters are optimized for operating zones relevant to the control policy. This framework bypasses the need for extensive high-quality historical database reflecting control data distributions that are rarely available in practice.
\par
Second, we formulate the performance loss that is necessary to compute the quality of the decisions and backpropagate the gradient with respect to the dynamics model parameters. Unlike in \cite{agrawal_learning_2020}, we do not assume that the system is differentiable. The only requirement is for the system state variables used in the system dynamics model to be observable. In addition, we introduce a hierarchical loss to inform the learning about the \acrshort{hvac} system structure and its impact on the objective value.
\par
Third, training is made robust to infeasibilities by relaxing constraints on the dynamics model output. Moreover, feasibility in the early stages of \acrshort{dfl} training is prompted by performing a warm-start with a model pre-trained on historical data. A small noise is added on the parameters after pre-training to get the model out of the local minimum while retaining a maximum of prior knowledge.
\par
The performance and relevance of the method is illustrated on the day-ahead \acrshort{hvac} scheduling of a realistic 15-zone building located in Denver, CO, US \cite{noauthor_commercial_nodate}. Furthermore, the method's robustness is evaluated under a distribution shift in the input parameters. Specifically, we analyze the model's performance metrics on a dataset representing an exceptionally hot year, a scenario likely to become increasingly common due to global warming.

\subsection{Outline}
Section \ref{sec: method} describes the problem class, explains gradient computations necessary to learn the parameters, the loss function design for non-differentiable black-box systems, and training robustness enabled by constraint relaxation. Afterwards, \autoref{sec: case study} presents the case study. In particular, sections \ref{sub: formulation} and \ref{sub: building performance loss} describe the specific optimization problem, and the loss function for the day-ahead \acrshort{hvac} scheduling, respectively. We report the results in section \ref{sub: results}. Section \ref{sec: discussion} discusses the results and limitations of the case study. Section \ref{sec: conclusion} concludes the paper and outlines future work directions.

\section{A Method for Simultaneous System Identification and Control}
\label{sec: method}

\begin{figure*}[th]
    \centering
    \includegraphics[scale=0.5]{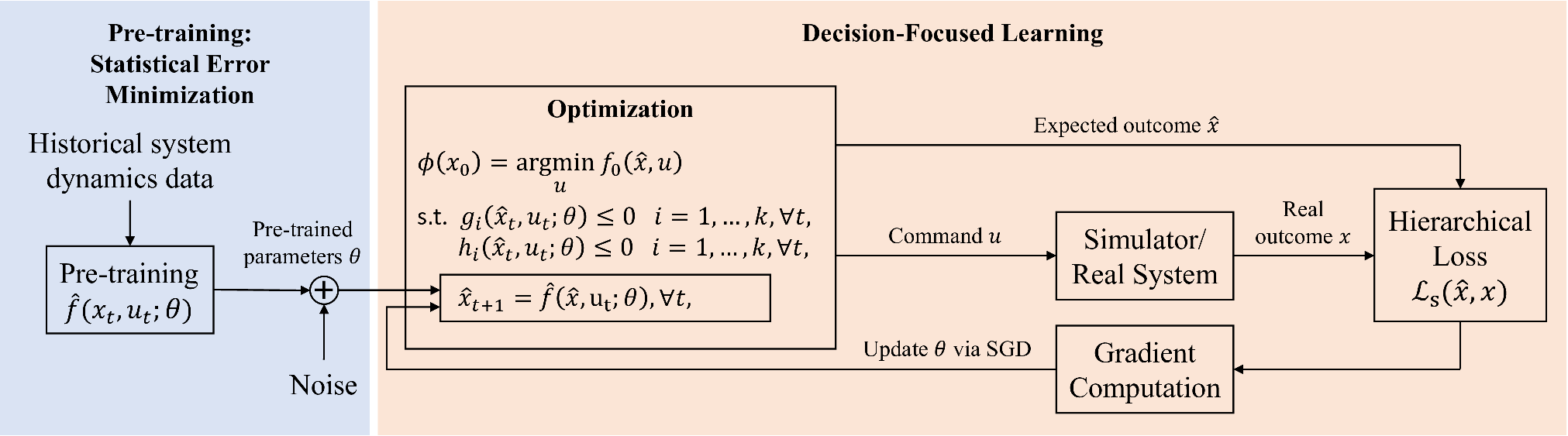}
    \caption{The proposed framework starts by pre-training the system dynamics model on historical data. Then, the pre-trained model parameters are fed into the convex optimization control policy where they keep being updated on representative scenarios to minimize the loss metric evaluating the mismatch between the expected and observed system states. }
    \label{fig: diagram}
    \Description{}
\end{figure*}
The proposed framework performs system identification and control simultaneously. To that end, we start by describing the addressed problem class, then come the explanation about the gradient computation needed for updating the parameters. Afterwards, we propose a \acrshort{dfl} supervised loss that bypasses the need for a differentiable system. Lastly, we address the infeasibility that may arise during training. \autoref{fig: diagram} provides an overview of the proposed method.

\subsection{Problem Class}
\label{sub: problem class}
Given a system with dynamics described by the unknown state transition function:
\begin{equation}
\label{eq: transition state function}
    x_{t+1} = f(x_t, u_t, w_t),\ \forall t,
\end{equation}
the goal is to learn the parameters $\theta$ of a proxy model $\hat f(\hat x_t, u_t; \theta)$ that represents the system dynamics, while keeping the formulation of the control policy $\phi$ convex \eqref{eq: problem class}. The known system state at a given time step $t \in \mathcal{T}$ is given by $x_t \in \R^n$, the expected state by $\hat x_t \in \R^n$, and the command or action by $u_t \in \R^m$. The true state transition function can be affected by some noise $w_t$. We do not make any assumption on $w_t$.\newline
A general formulation of the convex optimization control policy $\phi$ is given by:
\begin{flalign}
\label{eq: problem class}
    \phi(x_0) & = \argmin_u f_o(\hat x, u) \nonumber\\
    {\rm s.t.}\ & g_i(\hat x, u) \leq 0 \quad i = 1, ..., k,\ \forall t,\\
    & h_i(\hat x, u) = 0 \quad i = 1, ..., l,\ \forall t, \nonumber\\
    & \hat x_{t+1} = \hat f(\hat x_t, u_t; \theta),\ \forall t. \nonumber
\end{flalign}
To keep the problem convex, -the equality constraints $h_i$ and $\hat f$ must be affine with respect to the optimization variables $\hat x$ and $u$ whereas $g_i$ must be convex. The convexity of the problem is crucial for the gradient computation as explained in section \ref{sub: gradient computation}. In addition, convexity guarantees that the solution is the global optimum and offers tractable problems well handled by off-the-shelf solvers.
\par
Even though we focus on control optimization policy problems and state transition functions, the proposed framework can be used to learn any parameters of a convex optimization problem.

\subsection{Gradient Computation}
\label{sub: gradient computation}
Let us assume we have a scalar function $\mathcal{L}$ that quantifies the loss of performance (\textrm{i.e.}, the suboptimality) caused by the misestimated parameters $\theta$. The parameters $\theta$ can be learned while controlling the system with the policy $\phi$ by minimizing $\mathcal{L}$. This requires computing the gradient $\frac{\partial \mathcal{L}}{\partial \theta}$. Then, $\theta$ can be updated iteratively by any gradient-based optimization algorithm.
Misestimating $\theta$ leads to suboptimal commands $u$ that are responsible for the performance loss. By applying the chain rule, the gradient becomes:
\begin{equation}
\label{eq: performance loss chain rule}
    \frac{\partial \mathcal{L}}{\partial \theta} = \frac{\partial \mathcal{L}}{\partial u} \frac{\partial u}{\partial \theta}.
\end{equation}
The first factor accounts for the performance loss caused by a variation of the commands $u$.
The second factor creates the need for differentiating through the optimization problem. More specifically, the impact of a variation in the optimization parameters $\theta$ on the optimization output $u$ must be determined.
\par
An optimization problem can be considered as a mapping $O$ between its parameters and the optimal solution. Differentiating such a problem is challenging because of the complex and implicit mapping definition. Moreover, some classes of constrained programs, such as the linear programs, define piecewise-constant mappings in which the gradient is often null or nonexistent \cite{mandi_decision-focused_2023}.\newline
A general framework, named \textit{Cvxpylayers}, is able to differentiate conic programs \cite{agrawal_differentiating_2019}. Since any convex program can be recast as a conic program \cite{nemirovski_advances_2007}, \textit{Cvxpylayers} is very versatile. A conic program is of the form
\begin{flalign}
\label{eq: conic program}
    &\min_x c^Tx \\
    &s.t.\ b - Ax \in \mathcal{K}, \nonumber
\end{flalign}
where $(A, b, c) \in (\R^{m \times n}, \R^m, \R^n)$ are the parameters of the problem, $x \in \R^n$ is the primal decision variable, $\mathcal{K} \subseteq \R^m$ is a nonempty, closed, convex cone. Under the assumption that \eqref{eq: conic program} has a unique solution, its gradient can be computed. Solving a conic program can be done in three steps.
\renewcommand{\theenumi}{\roman{enumi}}
\begin{enumerate}
    \item The data parameters $(A, b, c)$ are used to build the corresponding skew-symmetric matrix $Q$ \cite{ye_onl-iteration_1994, odonoghue_conic_2016}.
    \item The self-homogeneous dual embedding problem --- which reformulates the KKT conditions of the original conic program into a single system of equations and incorporates the matrix $Q$ --- is solved by finding a root (\textrm{i.e.}, a zero) to its normalized residual map $s$ \cite{busseti_solution_2019}. The residual map is a function that quantifies the difference (or residual) between the current solutions of the primal and dual problems and the optimal solution. Therefore, it guides the search for the optimal solution, directing the optimization process.
    \item The solution $z$ is retrieved by $R$ and mapped to a valid solution of the original problem.
\end{enumerate}
Therefore, the mapping $O$ can be seen as the composition of three submappings $R \circ s \circ Q$. By the chain rule, we have
\begin{equation}
    {\rm D} O(A,b,c) = {\rm D} R(z)\ {\rm D} s(Q) \ {\rm D} Q(A,b,c)
\end{equation}
with $\rm D$, the derivative operator. The derivative of $Q$ is straightforward since $A$, $b$, and $c$ appear explicitly in its formulation. The derivative of the root-finding problem $s$ is obtained via implicit differentiation. The derivative of the retriever $R$ relies on the derivative of the euclidean projection of the self-homogeneous dual embedding solution $z$ onto the feasible cone $\mathcal{K}^\star$ associated with the dual of the original conic problem \eqref{eq: conic program} \cite{agrawal_differentiating_2019}.

\subsection{Performance Loss for Non-Differentiable System}
\label{sub: performance loss}
As established by \eqref{eq: performance loss chain rule}, the gradient of the performance loss with respect to the parameters $\theta$ must be defined and exist. The ideal loss prescribed in the literature is the regret $\mathcal{L}_r$ \eqref{eq: regret} \cite{adam_n_elmachtoub_smart_2017, mandi_decision-focused_2023}. The regret assesses the objective value loss at optimality (marked by $^*$) caused by the error between the actual $\theta$ and its estimator $\hat \theta$:
\begin{equation}
\label{eq: regret}
    \mathcal{L}_r = f_o^*(\hat \theta) - f_o^*(\theta).
\end{equation}
However, the definition of regret assumes the actual value of $\theta$ is known. Other loss functions have considered the transition state function $f$ \eqref{eq: transition state function} as fully known and differentiable \cite{agrawal_learning_2020}.
\par
Many physical systems must be considered as black boxes, and thus non-differentiable, because of their inherent complexity and the unbearable cost to build an accurate system model. The gradient of black-box model can be estimated via simultaneous perturbation stochastic approximation \cite{chau_overview_2015}. However, it is unpractical if the system environment conditions are evolving (e.g. the weather) making the repetition of different commands in the same conditions impossible.\newline
Consequently, we design a novel performance loss $\mathcal{L}_{\rm DFL}$ that does not assume any knowledge about the system:
\begin{equation}
\label{eq: supervized DFL}
    \mathcal{L}_{\rm DFL} = \mathcal{L}_{s}(\hat x, x),
\end{equation}
where $\mathcal{L}_s$ refers to any supervized loss such as the Mean Squared Error (MSE) loss or the Mean Absolute Error (MAE) loss. The loss $\mathcal{L}_{\rm DFL}$ computes a statistical metric on the error between the expected system state $\hat x$, as predicted by the optimization model, and the actual observed realisation $x$. Similar to supervized learning, there is a target value, $x$, to which no gradient is attached. We refer to  $\mathcal{L}_{\rm DFL}$ as the supervized \acrshort{dfl} loss. Unlike traditional methods that rely solely on historical data and remain agnostic on the downstream control policy, our approach benefits of more informed sampling of the input space $u$. This allows the state-transition model to allocate its modeling resources more effectively. Consequently, input regions that are critical to the control policy are modeled with greater precision, enabling even simple models to achieve high accuracy.

Interestingly, the regret $\mathcal{L}_r$ and the supervized \acrshort{dfl} losses $\mathcal{L}_{\rm DFL}$ do not pursue the same objective. On the one hand, the regret $\mathcal{L}_r$ trains the parameter $\hat \theta$ to bring the obtained objective value as close as possible to the objective value obtained with perfect knowledge of $\theta$. On the other hand, the supervized \acrshort{dfl} loss aims at improving the ability of the state transition proxy model $\hat f$ to match the true outcome. Notably, all performance losses would be the same if $\hat f$ is an error-free approximation of $f$.

\subsection{Robustness to infeasibility}
\label{sub: robustness to infeasibility}
The updates of the constraint parameters by the SGD may lead the optimization program to become infeasible. A small number of infeasible optimization problems over a full training epoch may not be an issue if the remaining solvable samples update the parameters in a way which restores feasibility. However, such an approach lacks robustness. Therefore, we relax the constraints bounding the output of the system state model and formulate them as a quadratic regularization term in the objective function. To that end, the state model output is associated with a target value $X^{\rm g}$. The target value is the desired state for the system and is chosen by the control policy designer. For example, the target value for the indoor temperature might be set to 21°C (70°F).
A weight $w$ defines the importance of meeting the target value. In addition to preventing infeasibility, the quadratic regularization term can transform linear programs into strongly convex quadratic programs, which, unlike linear programs, provide a non-null (and thus informative) gradient:
\begin{flalign}
\label{eq: relaxed problem}
    \phi(x) & = \argmin_u f_o(x, u) + w * (x - X^{\rm g})^2 \nonumber\\
    {\rm s.t.} \ & f_i(u) \leq 0 \quad i = 1, ..., k\ \forall t,\\
    & h_i(u) = 0 \quad i = 1, ..., l\ \forall t, \nonumber\\
    & \hat x_{t+1} = \hat f(\hat x_t, u_t; \theta),\ \forall t. \nonumber
\end{flalign}
In addition, we propose a pre-training stage where the dynamics model is trained on historical data. It provides the policy with an initial estimate of the dynamics model parameters, which improves feasibility in the early stages of training. However, minimizing a task-agnostic loss on historical data may lead the pre-training to converge to a local minimum of the supervised DFL loss. Escaping this local minimum might require using a large step size in the gradient descent algorithm, which could harm training convergence. To address this, we apply calibrated Gaussian noise to the pre-trained parameters. The intensity of the noise must be carefully selected to retain as much pre-training information as possible while helping the model escape the local minimum.

\section{Building Thermodynamics Modeling and Day-ahead Control}
\label{sec: case study}
We demonstrate the effectiveness of our method on the hourly day-ahead \acrshort{hvac} scheduling of a three-floor medium office building with 15 conditioned zones. We aim at learning the parameters of a multi-zone \acrshort{rc} formulation modeling the building thermodynamics, while optimizing \acrshort{hvac} scheduling. In this section, we start by describing the building. The formulation of the optimization problem follows. Then, the data and their preprocessing are presented. Afterwards, we report and analyze the results of the proposed approach before comparing it to the conventional two-stage \acrshort{ito} performances. Finally, we assess the robustness of the model to distribution shift of the input data.

\subsection{Building Description}
\label{sub: building template}
\begin{figure*}
    \centering
    \includegraphics[width=0.65\textwidth]{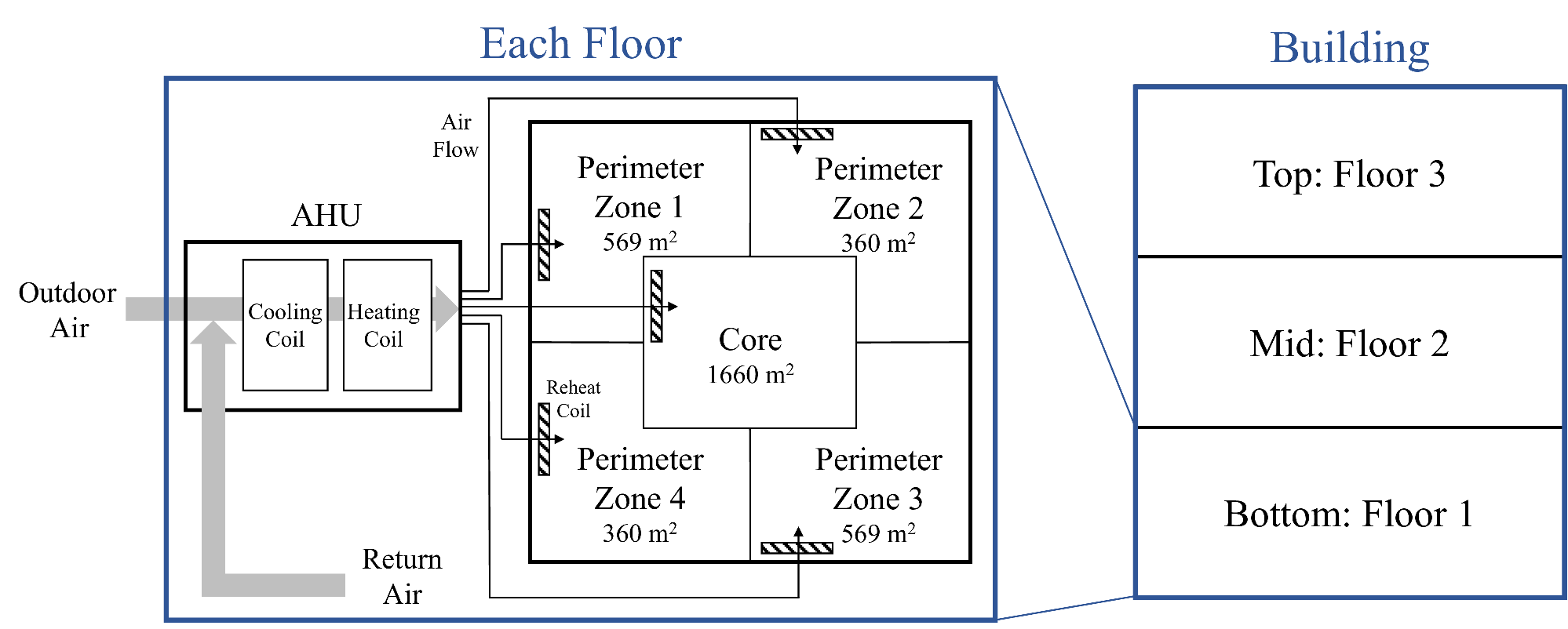}
    \caption{The building is made of three similar floors. Each floor is equipped with its own variable air volume air handling unit responsible for the thermal comfort of the five zones.}
    \Description{}
    \label{fig: AHU design}
\end{figure*}

The building model is provided by the U.S. Department of Energy as part of EnergyPlus \cite{crawley_energyplus_2001}. EnergyPlus is a state-of-the-art physics-based high-fidelity simulator for buildings. The building is an office building of 4928~m$^2$ (53,628~ft$^2$) assumed to be located in Denver, CO. It comprises three floors of six zones each, only five being thermally controlled. The sixth zone is a plenum that contains the ducts transporting the conditioned air. Each floor is equipped with an Air Handling Unit (AHU). An AHU is an equipment that centrally conditions the air (either cooling or heating) before distribution to the zones via the duct network. The AHU is also responsible for the ventilation of the zones. The system in this building is a variable air volume system with reheat. The air flow rate reaching each zone can be modulated individually and, if necessary, the air can be heated just before entering the zone. \autoref{fig: AHU design} provides a layout of the building floor and the AHU architecture. We make two assumptions with respect to the original template:
\renewcommand{\theenumi}{\roman{enumi}}
\begin{enumerate}
    \item we replace the gas-fired heating coil of the AHU with an efficiency of 81~\% by an electrical coil with an efficiency of 100~\%;
    \item we divide the AHU electricity consumption proportionally to the air mass flow rate of each zone served by the AHU, while the reheat coil electricity consumption is associated with the zone it supplies.
\end{enumerate}
The building features a conventional occupancy schedule for offices; most of the activity takes place during weekdays from 7am to 6pm.

\subsection{Control Formulation}
\label{sub: formulation}

The control problem aims to find the temperature profile for each zone $\tau_{t, z}^{\rm in}$ that minimizes the electricity cost of \acrshort{hvac} while ensuring thermal comfort \eqref{eq: obj fct}. The set of zones and time steps are $\mathcal{Z}$ (with cardinality $Z$, index $z$) and $\mathcal{T}$ (with cardinality $T$, index $t$), respectively. The electricity price consists of two terms. The first term (\textit{i}) is the demand charge. The demand charge is proportional to the daily power consumption peak $p^{\rm d}$ and cost $\lambda^{\rm d}$. The demand charge reflects the cost of network infrastructure to supply the requested power. The second term (\textit{ii}) is the cost for the energy consumed over the time step $p_t^{\rm i}$ at price $\lambda_t^{\rm i}$ following the time-of-use tariff. Finally, the regularization term (\textit{iii}) guarantees the problem feasibility, even during \acrshort{dfl}. The regularization term penalizes the deviation of the indoor temperature $\tau_{t, z}^{\rm in}$ from the target $T_{t, z}^{\rm tgt}$ with a quadratic term. The weight $w_{t, z}$ adjusts the penalty incurred for thermal discomfort (\textrm{i.e.}, the indoor temperature deviation from the target value) through the zones and time.
\begin{flalign}
    \label{eq: obj fct}
    \min_{\tau^{\rm in}} \quad \underbrace{\vphantom{\sum_{t=0}^{T}}p^{\rm d} \lambda^{\rm d}}_\text{(\textit{i})} + \sum_{t=0}^{T} \left( \underbrace{\vphantom{\sum_{t=0}^{T}} p_t^{\rm i} \lambda_t^{\rm i} }_\text{(\textit{ii})} + \underbrace{ \sum_{z=0}^{Z} w_{t, z} (\tau_{t, z}^{\rm in} - T_{t, z}^{\rm tgt})^2}_\text{(\textit{iii})} \right) \Delta t
\end{flalign}

The \acrshort{rc} model in \eqref{eq: rc model} accounts for  multi-zonal building dynamics, where matrix $\alpha \in \R^{Z\times Z}$ represents inter-zonal heat transfer,  vectors $R \in \R^{Z}$ and $C \in \R^{Z}$ are zonal lumped resistances and capacitances, and vectors $\eta^{\rm h} \in \R^{Z}$ and $\eta^{\rm c}\in \R^{Z}$ are heating and cooling efficiencies, which include thermal losses of the duct system. For the 15-zone building, the \acrshort{rc} model features 265 parameters. The parameters $\theta = (\alpha, \eta^{\rm c}, \eta^{\rm h}, R, C)$ will be learned while scheduling the \acrshort{hvac} system.

\begin{flalign}
    \label{eq: rc model}
    \tau_{t+1}^{\rm in} = \Delta t \left( \alpha \tau_{t}^{\rm in} + \frac{\left(\eta^{\rm h} p_{t}^{\rm h} - \eta^{\rm c} p_{t}^{\rm c}\right)}{C}  + \frac{(\tau_{t}^{\rm amb} - \tau_{t}^{\rm in})}{RC} \right) \ \forall t
\end{flalign}
The initial conditions are given by: 
\begin{equation}
    \tau_{0, z}^{\rm in} = T_{0, z}^{\rm in} \quad \forall z.
\end{equation}
For each zone, the \acrshort{hvac} capacities for cooling $\overline{P}_{t, z}^{\rm c}$ and heating $\overline{P}_{t, z}^{\rm h}$ are determined based on historical data. The constraints are imposed at the zonal \eqref{eq: zonal hvac capacity} and floor \eqref{eq: floor hvac capacity} levels. The floor level constraint is essential to capture the maximum consumption of each AHU while the zonal constraint limits the amount of energy that can be dedicated to a specific zone. The set of floor is $\mathcal{F}$ with cardinality $F$, and index $f$. Each floor $f$ is a set containing the zones $z$ of that floor.
\begin{flalign}
    \label{eq: zonal hvac capacity}
    p_{t, z}^{\rm c} \leq \overline{P}_{t, z}^{\rm c},&\ p_{t, z}^{\rm h}  \leq \overline{P}_{t, z}^{\rm h} \quad \forall t, z\\
    \label{eq: floor hvac capacity}
    \sum_{z \in f} p_{t, z}^{\rm c} \leq \overline{P}_{t, f}^{\rm c},&\ \sum_{z \in f} p_{t, z}^{\rm h}  \leq \overline{P}_{t, f}^{\rm h} \quad \forall t, f
\end{flalign}

We define the \acrshort{hvac} electricity power as the sum of the cooling and heating electricity powers \eqref{eq: hvac power}. Eq. \eqref{eq: energy balance} maintains the energy balance. In this reduced form, the \acrshort{hvac} electricity must be purchased from the grid.
\begin{flalign}
\label{eq: hvac power}
    p_{t, z}^{\rm hvac} &= \ p_{t, z}^{\rm h} + \ p_{t, z}^{\rm c} \quad \forall t, z\\
\label{eq: energy balance}
    \sum_{z=0}^{Z} p_{t, z}^{\rm hvac} &= p_t^{\rm i} \quad \forall t
\end{flalign}
Constraint \eqref{eq: peak demand} defines the power peak demand as being greater than all power demand. Because the peak demand directly increases the electricity cost, this constraint is equivalent to $p^{\rm d} = \max {\{p_t^{\rm i}: t\in T\}}$, but it avoids making the problem bilevel.
\begin{equation}
\label{eq: peak demand}
    p^{\rm d} \geq p_t^{\rm i} \quad \forall t
\end{equation}
Finally, the power imported from the grid must be inferior to the line capacity $\overline{P}^{\rm i}$. This is equivalent to imposing the peak power demand to be inferior to $\overline{P}^{\rm i}$ \eqref{eq: line capacity}. All power levels must be positive \eqref{eq: positivity}.
\begin{align}
\label{eq: line capacity}
    p^{\rm d} &\leq \overline{P}^{\rm i} \\
\label{eq: positivity}
    p_{t, z}^{\rm c},\ p_{t, z}^{\rm h},\ p_t^{\rm i} &\geq 0 \qquad \forall t, z
\end{align}

\subsection{Datasets and Clustering}
\label{sub: dataset}

\begin{figure}[t]
    \centering
    \includegraphics[scale=0.2]{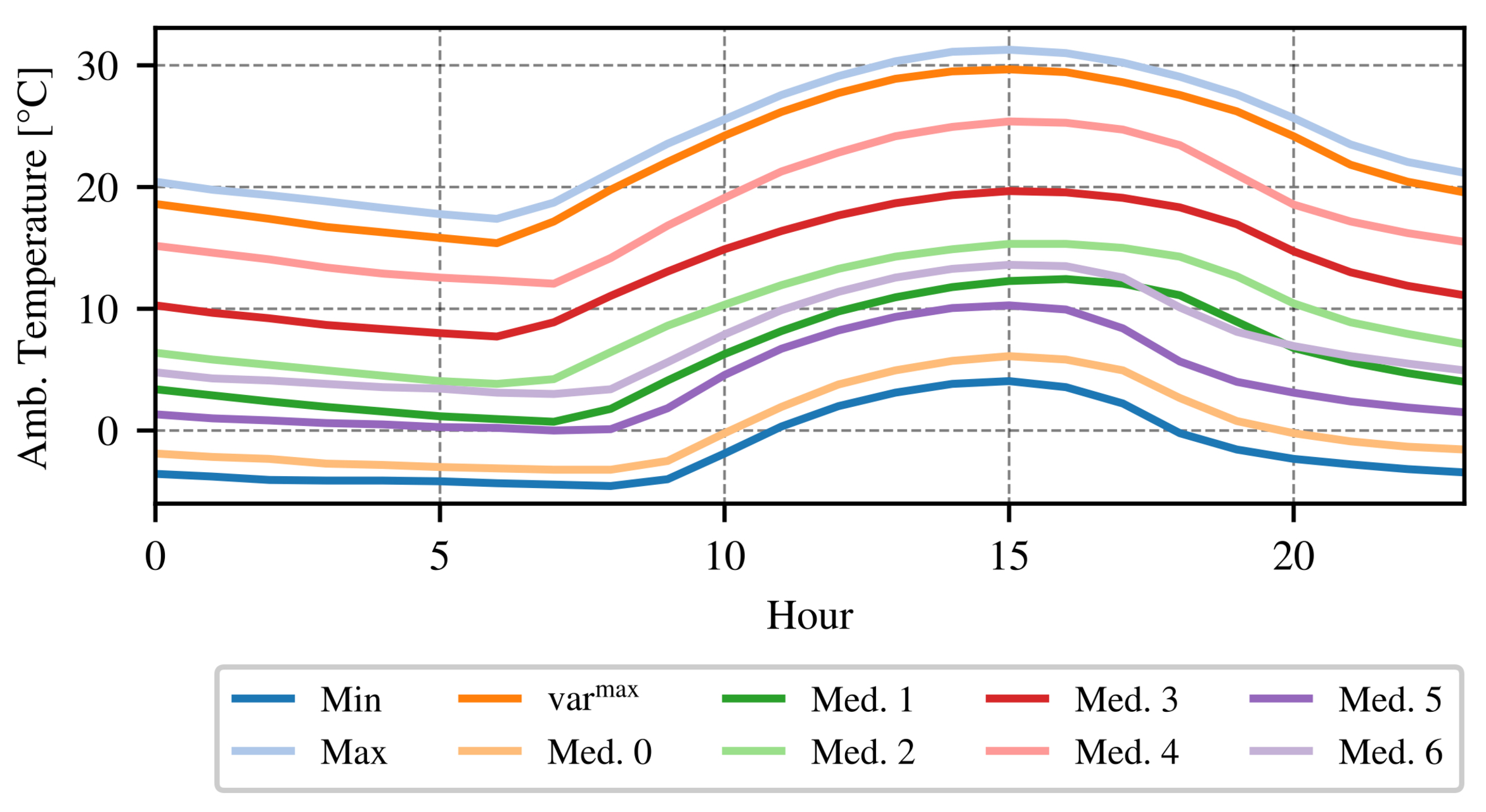}
    \caption{Ambient temperature profiles of the ten medoids. The three first labels are extreme scenarios.}
    \Description{}
    \label{fig: medoids}
\end{figure}

\begin{figure}[t]
    \centering
    \includegraphics[scale=0.8]{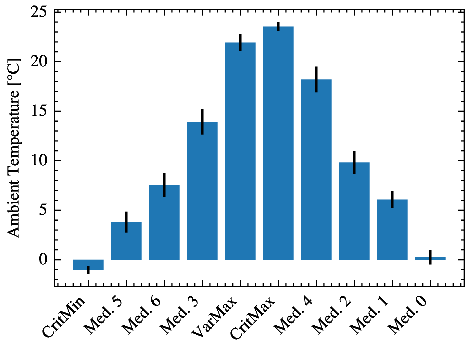}
    \caption{Mean and standard deviation of the medoids. They are ordered to form a smooth cycle.}
    \Description{}
    \label{fig: clusters}
\end{figure}

To simulate building thermodynamics, two typical meteorological datasets are used. A typical meteorological dataset is a representative year of hourly weather data for a specific location, compiled from long-term observations. The first dataset is used as input to generate one year of historical data \textit{without} any optimal scheduling. The second dataset provides weather scenarios for the scheduling optimization. Since optimizing and simulating the daily schedule for the 365 days of the year is burdensome, we perform a k-medoid clustering on the 365 days of weather data in the second dataset. The k-medoid algorithm is preferred to the k-mean algorithm to avoid the creation of fictitious average data. We focus on the ambient temperature because it is the only weather measurement necessary to the \acrshort{rc} model \eqref{eq: rc model}. First, we identify three extreme days: the coldest, the hottest, and the day with the highest temperature variance. These three days are three fixed medoids of our medoid search. Then, we look for seven additional medoids to obtain the best partition of the space. The resulting temperature profiles of the k-medoid algorithm are given in \autoref{fig: medoids}. The cluster means and standard deviations are reported in \autoref{fig: clusters}.

% Diagram figure
\begin{figure*}[!t]
    \centering
    \includegraphics[scale=0.5]{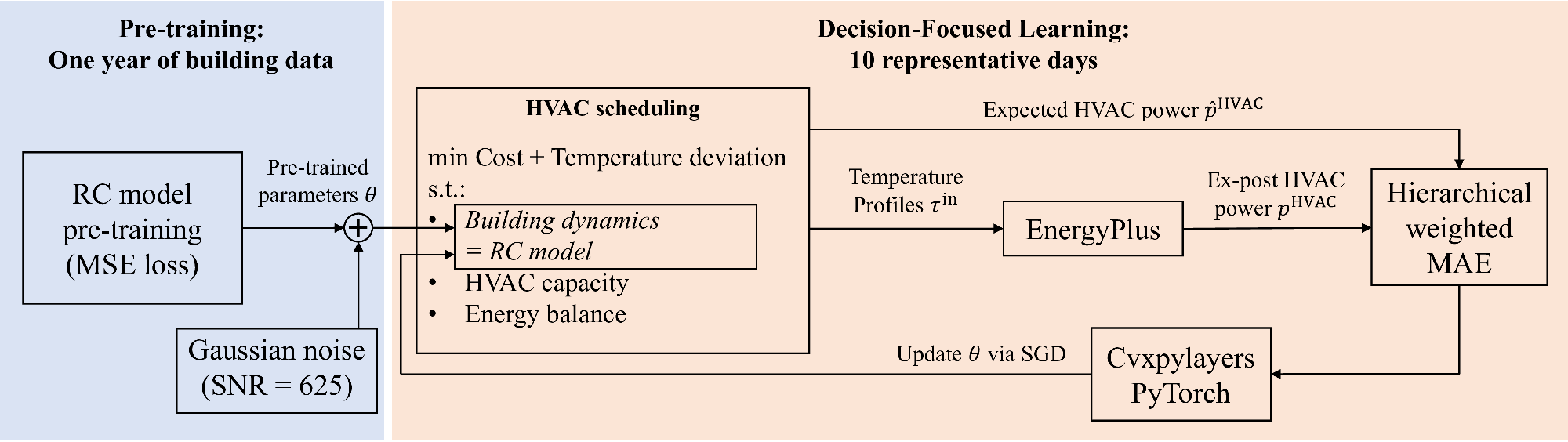}
    \caption{The building \acrshort{rc} model is pre-trained on one year of historical data before a small noise is applied on the parameters. They are then updated to perform the best day-ahead \acrshort{hvac} scheduling over 10 representative days with respect to the task-aware MAE loss.}
    \Description{}
    \label{fig: hvac diagram}
\end{figure*}

\subsection{Hierarchical Performance Loss}
\label{sub: building performance loss}
As seen in section \ref{sub: performance loss}, we can define any supervized \acrshort{dfl} loss for learning the parameters $\theta = (\alpha, \eta^{\rm c}, \eta^{\rm h}, R, C)$ over the ten daily scenarios selected. In this specific case, the command is the indoor temperature profile for each zone $\tau_{t, z}^{\rm in}$ since it is the input of the thermostat. Therefore, the DFL loss focuses on the difference between the expected and the ex-post \acrshort{hvac} power consumptions. The underlying goal is to accurately capture the impact of \acrshort{hvac} power prediction errors on the objective value. Because the \acrshort{hvac} power $p^{\rm hvac}$ appears linearly in the objective function, we use a linear performance loss, the MAE. Furthermore, we weight the loss at each time step with the all-inclusive price coefficient $\lambda_t^{\rm i,d}$ that adds the peak demand price $\lambda_t^{\rm d}$ to the energy prices $\lambda_t^{\rm i}$for the time step with the highest ex-post power. Otherwise, $\lambda_t^{\rm i,d}$ and $\lambda_t^{\rm i}$ are equal. Finally, we leverage hierarchical loss. By order of importance, we minimize the error for (\textrm{i}) the whole building to obtain an accurate expectation of the electricity bill; (\textrm{ii}) each AHU which means for the sum of the five zones at each floor $f$; (\textrm{iii}) each zone $z$.
\newline
The final loss function is given by \eqref{eq: case study loss function}. The number of zones at each loss level is used to weight the importance. The building encompasses 15 zones ($w_b = 15$) while each floor features five zones ($w_f = 5 \ \forall f$).
\begin{align}
\label{eq: case study loss function}
    L_{\rm DFL} = &\frac{1}{T} \sum_{t=0}^T \lambda_t^{\rm i,d}
    \left( \vphantom{\sum_{z=0}^Z {\rm MAE}}
    w_b \cdot {\rm MAE} \left( \hat p_{t, b}^{\rm hvac}, p_{t, b}^{\rm hvac} \right) \right. \nonumber \\
    &+ w_f \sum_{f=0}^F {\rm MAE} \left( \hat p_{t, f}^{\rm hvac}, p_{t, f}^{\rm hvac} \right) \\
    &+ \left. \sum_{z=0}^Z {\rm MAE} \left(\hat p_{t, z}^{\rm hvac}, p_{t, z}^{\rm hvac} \right) \right) \nonumber
\end{align}
where $\hat p_{t, f}^{\rm hvac} = \sum_{z \in f} p_{t, z}^{\rm hvac}$ is the \acrshort{hvac} power of floor $f$ and $\hat p_{t, b}^{\rm hvac} = \sum_{z = 0}^Z p_{t, z}^{\rm hvac}$ is the \acrshort{hvac} power of the whole building.

\begin{figure*}[th]
    \centering
    \subfigure[]{\includegraphics[width=0.3\textwidth]{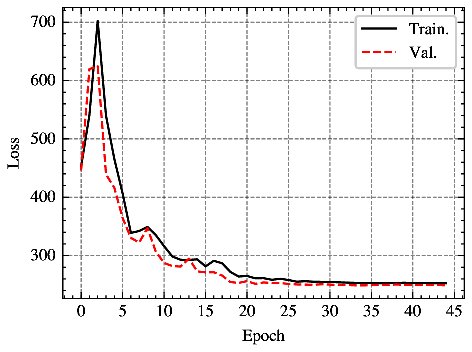}}
    \quad  % Add space between subplots
    \subfigure[]{\includegraphics[width=0.3\textwidth]{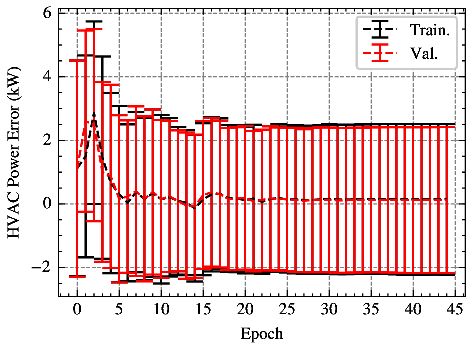}}
    \quad  % Add space between subplots
    \subfigure[]{\includegraphics[width=0.3\textwidth, height=0.225\textwidth]{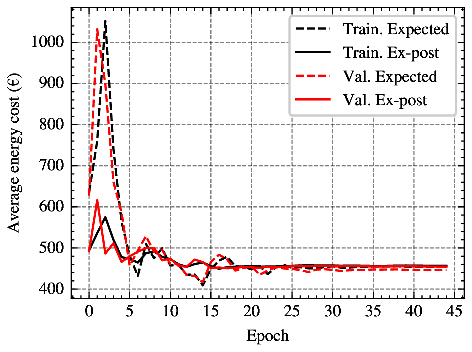}}
    \caption{(a) The \acrshort{dfl} supervised loss (hierarchical weighted MAE) converges at much lower levels than initially; (b) the error mean converges towards zero and the standard deviation reduces; (c) the expected cost becomes a more accurate prediction of the ex-post cost while diminishing.}
    \Description{}
    \label{fig: training performances}
\end{figure*}

\subsection{Results}
\label{sub: results}
Our goal is to learn the parameters $\theta = (\alpha, \eta^{\rm c}, \eta^{\rm h}, R, C)$ that define the building thermodynamics \eqref{eq: rc model}. We initialize the algorithm by the pre-training. During this warm-start period, the parameters are trained over the 365 days of the first typical weather file using an \acrshort{mse} loss. Then, an element-wise Gaussian noise $N \sim \mathcal{N}(0, \theta_i/25)$,  which corresponds to a Signal-to-Noise Ratio (SNR) of 625 (\textrm{i.e.}, $25^2$), is applied independently on each element of the parameter vector. This operation aims at getting the \acrshort{rc} model out of the pre-training local minimum while retaining a maximum of information.
\par
Afterwards, we enter the \acrshort{dfl} stage. The \acrshort{dfl} training set is made of ten representative days obtained via clustering. In addition, we sample randomly in each cluster two days to form a validation set and a test set of ten days. We consider solving and simulating the solution once for each of the ten training problems as a training epoch of ten samples. Each problem is solved using the \textit{ECOS} solver since it was reported to be the best performing solver for conic \acrshort{dfl} \cite{wahdany_more_2023}. \textit{Cvxpylayer} computes the gradient of the optimization problem and \textit{PyTorch} performs the backpropagation. As soon as a sample (\textrm{i.e.}, a day) has been solved and simulated, the parameters are updated (\textrm{i.e.}, pure Stochastic Gradient Descent (SGD)). To facilitate learning, we order the days to form a smooth cycle when looping over the epochs as depicted by \autoref{fig: clusters}. We initialize \textit{Adam}, a gradient-based optimization algorithm, with a learning rate at 0.001 and a polynomial decay of $\gamma = 0.9$ \cite{kingma_adam_2017}. We set the maximum number of epochs to 50 with an early stopping featuring a 10-epoch patience. \autoref{fig: hvac diagram} represents the whole process.
% Table \ref{tab: learning parameters} gathers the parameter values. 
% \begin{table}[ht]
%     \centering
%     \begin{tabular}{l r}
%     \toprule
%         SNR & 900 \\
%         Learning rate & $10^{-3}$ \\
%         Exponential decay $\gamma$ & 0.9 \\
%         Training set size & 10 \\
%         Validation set size & 10 \\
%         Maximum number of epochs & 50 \\
%         Patience & 15 \\
%     \bottomrule
%     \end{tabular}
%     \caption{Learning parameters}
%     \label{tab: learning parameters}
% \end{table}

\begin{figure}[ht]
    \centering
    \includegraphics[scale=0.8]{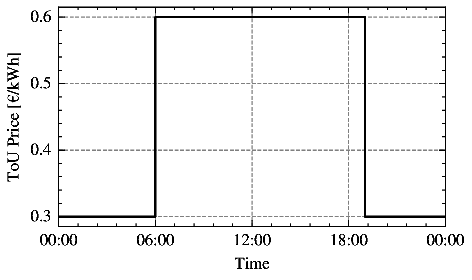}
    \caption{Time-of-use tariff with an appreciation of electricity from 6am to 7pm.}
    \Description{}
    \label{fig: time-of-use tariff}
\end{figure}

The time-of-use tariff is a rectangular function. From 7pm to 6am (excluded) the price is 0.3~\euro/kWh. It rises to 0.6~\euro/kWh from 6am to 7pm (excluded) as depicted by \autoref{fig: time-of-use tariff}. The increase highlights the higher demand during the day.
% around 12:30am reflecting the higher electricity prices during the day. The demand charge is set to 0.4~\euro/kW.

The performances of the \acrshort{dfl} training are presented in \autoref{fig: training performances}. Overall, the training and validation metrics follow very similar trends. This suggests that ten medoids are sufficient for the model to generalize effectively. The first plot displays the evolution of the loss over the training epochs. The training is interrupted after 44 epochs by the early stopping because the minimum value for the validation loss is reached at the 34\textsuperscript{th} epoch. Both the training and validation losses tend to decrease until the 34\textsuperscript{th} epoch before leveling off. At epoch 34, the training and validation losses stand at 252 and 248, respectively.\newline
The center plot (figure \ref{fig: training performances}.b) shows the evolution of the error, assuming a Gaussian distribution. After about five epochs, the error mean converges around 0 with a standard deviation at about 2.5~kW. Then, the error mean ripples slightly until the 25\textsuperscript{th} epoch where it stabilizes at 0.15 with a standard deviation of 2.35. The training and validation sets exhibit very similar trends.\newline
The last plot displays the evolution of the expected and ex-post electricity costs for the training and validation sets. The expected and ex-post costs are very volatile over the first half of the training. Over the second half, the expected and ex-post costs level off at about 450~\euro\ and 455~\euro, respectively. The bill prediction of the day-ahead planning is therefore accurate. It is worth noting that the training converges at the minimum ex-post cost over all epochs.\newline
In conclusion, when surveying simultaneously the three metrics, the set of parameters obtained at the end of the 34\textsuperscript{th} epoch offers the highest performance. 
\par
In comparison, the \acrshort{ito} model performs extremely poorly. The \acrshort{rc} parameters of the \acrshort{ito} model were found by minimization of the \acrshort{mse}. Nevertheless, the \acrshort{mse} reported in Table \ref{tab: ITO vs DFL} indicate clearly that the \acrshort{dfl} model at the 34\textsuperscript{th} epoch is better for each data set: training, validation, and test. The two models show very consistent metrics across the three data sets.\newline
The hierarchical loss of the \acrshort{ito} test set stands at 652, more than two and a half times the value of the \acrshort{dfl} model (253). However, the MAE is slightly lower for the \acrshort{ito} model, rising from 2.66 for the \acrshort{ito} model to 2.94 for the \acrshort{dfl} model. This can be explained by the hierarchical loss that does not aim at minimizing the prediction error at the zonal level. The test set average error of the \acrshort{ito} model stands at -1.81~kW with a standard deviation of 2.39~kW. Since, on average, the expected power is much lower than the ex-post power, the electricity cost is underestimated. The expected cost is 85~\euro\ for an ex-post cost at 474~\euro. It represents a prediction error of 389~\euro. In contrast, the \acrshort{dfl} model achieve a prediction error of 16~\euro. Very interestingly, the ex-post cost of the \acrshort{dfl} model (468~\euro) is lower than for the \acrshort{ito} model (474~\euro). The foremost hindrance to \acrshort{dfl} is the heavy computational burden associated with training. On a personal MacBook Pro, equipped with an Apple M1 Pro chip and 32 GB of memory, the \acrshort{dfl} training took 7 hours compared to 16 minutes for the conventional supervized learning.
\begin{table}[ht]
\centering
\begin{tabular}{@{}lllllll@{}}
\toprule
 & \multicolumn{3}{c}{ITO} & \multicolumn{3}{c}{DFL} \\
                                     & Train. & Val. & Test & Train. & Val. & Test \\ \midrule
Hierarchical loss                    & 655    & 646    & 652   & 252    & 248    & 253  \\
MAE (kW)                             & 2.68   & 2.62   & 2.66  & 2.93   & 2.87   & 2.94 \\
MSE (kW\textsuperscript{2})          & 16.8   & 16.6   & 16.7  & 15.4   & 14.6   & 15.6 \\
Error mean (kW)                      & -1.85  & -1.86  & -1.81 & 0.15   & 0.12   & 0.08 \\
Error std (kW)                       & 2.34   & 2.28   & 2.39  & 2.35   & 2.27   & 2.38 \\
Expected cost (\euro) & 82     & 75     & 85    & 454    & 446    & 452  \\
Ex-post cost (\euro)  & 478    & 474    & 474   & 456    & 455    & 468  \\
Cost error (\euro)    & 395    & 398    & 389   & 2      & 8      & 16 \\
Training Time   &   \multicolumn{2}{l}{16~min} &    & \multicolumn{2}{l}{7h02min} &   \\
\bottomrule
\end{tabular}
\caption{Performances of the Identify-Then-Optimize (ITO) model versus the Decision-Focused Learning (DFL) model at epoch 34.}
\label{tab: ITO vs DFL}
\end{table}

\begin{figure*}[!ht]
    \centering
    \begin{tabular}{c c}
        \raisebox{12em}{(a)} & \includegraphics[width=0.9\textwidth]{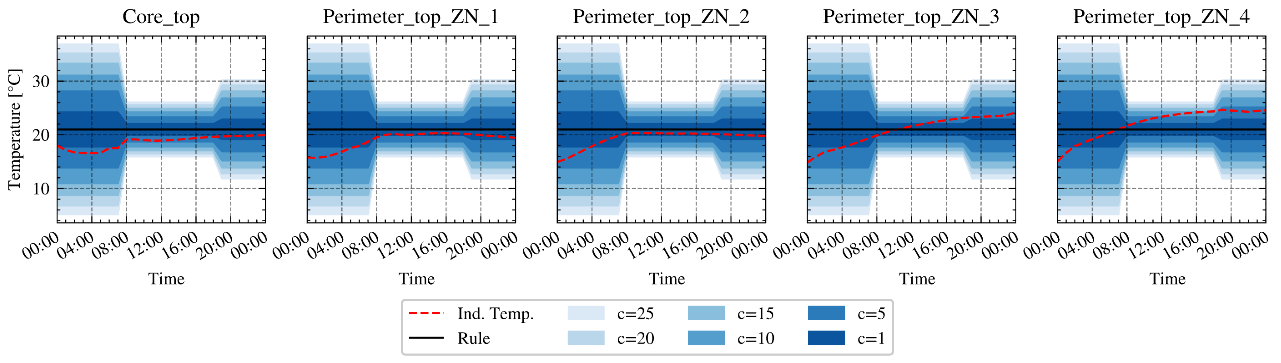}
    \\
        \raisebox{11em}{(b)} & \includegraphics[width=0.9\textwidth]{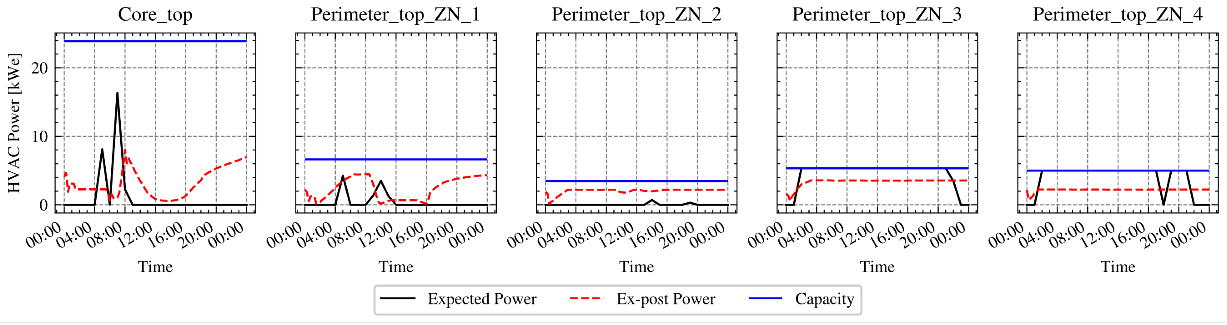}
    \\
        \raisebox{11em}{(c)} & \includegraphics[width=0.9\textwidth]{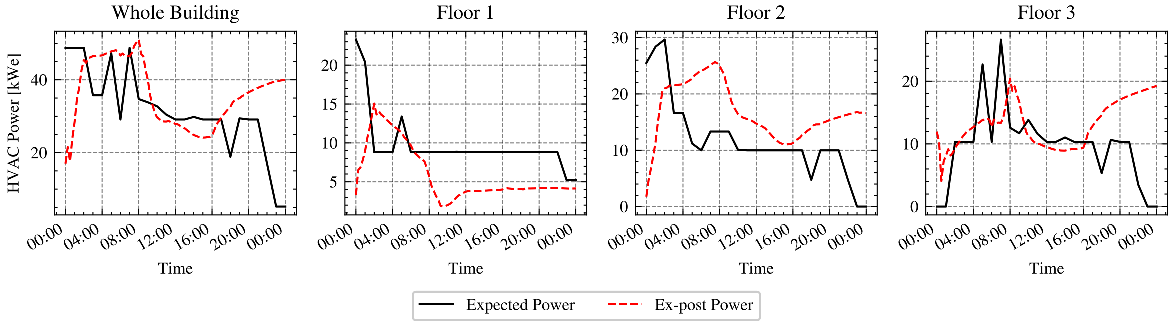}
    \end{tabular}
    \caption{Scheduling for the coldest day obtained by the \acrshort{dfl} \acrshort{rc} model at epoch 34. Subfigures are (a) the zonal indoor temperature of the top floor; (b) the zonal \acrshort{hvac} electric power of the top floor, and (c) the aggregated \acrshort{hvac} power for the whole building and each floor. The blue shades indicate the temperature deviations causing an objective value penalty lower than a given threshold. Temperature control is more loose during the night and evening because of the worker absence. The mismatch between the expected and ex-post \acrshort{hvac} powers in (b) can be explained by the intrinsic \acrshort{rc} model limitations and the task-aware loss which gives little importance to accurate zonal predictions.}
    \Description{}
    \label{fig: temperature and power}
\end{figure*}
\par
The scheduling of the building lower floor on the coldest day is shown in \autoref{fig: temperature and power}. The upper part displays the temperature profiles within the five zones. The blue shades correspond to a given penalty per hour for deviating from the 21°C target. The discomfort penalty is the lowest during the night, from midnight to 7am, and the highest during the conventional working hours, reflecting the higher occupancy.\newline
Overall, and aligned with the discomfort penalty, the temperature profiles stay close to the target of 21°C during the day, with greater deviations occurring in the evening and night. Still the deviations remain within admissible ranges. The electric power scheduling of the \acrshort{hvac} is shown in the lower two-thirds figure.
The zonal electric power scheduling in subfigure \ref{fig: temperature and power}.b is quite inaccurate. This can be explained by the hierarchical loss, which focuses primarily on having a precise day-ahead scheduling of the whole building. This is depicted by subfigure \ref{fig: temperature and power}.c where the power trends are much better for the floor level and the entire building. At the building level, the attempt to shift the \acrshort{hvac} consumption towards cheap electricity stands out. Indeed, the \acrshort{hvac} consumption is high during the night, from 2am to 8am, and peaks again towards the end of the day, when the price of electricity goes down. The \acrshort{rc} model is a linear approximation of the dynamics considering only the previous time step. Moreover, the outdoor temperature is the only weather input. Consequently, the limited modeling capability and input of the \acrshort{rc} model, along with the complexity of the building's thermodynamics, are responsible for the remaining inaccuracies. The real-time adjustment to the model inaccuracies and the changing conditions should be handled by a real-time controller.

\subsection{Ambient Temperature Shift}

\begin{figure}[ht]
    \centering
    \includegraphics[width=0.8\linewidth]{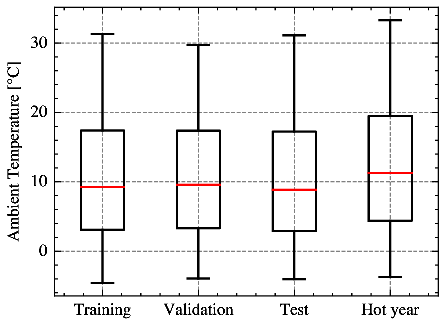}
    \caption{Ambient temperature distribution of each data set. Whiskers represent minimum and maximum temperatures.}
    \Description{}
    \label{fig: box plot}
\end{figure}

We investigate the robustness of the proposed framework to the input distribution shift. In particular, we assess the model response to a hot year. We build the hot year by selecting the hottest sample from each cluster. For the sample associated with the hottest cluster, we add two Celsius degrees on top of the hottest day, assuming record-breaking temperatures. \autoref{fig: box plot} compares the ambient temperature distribution of the four data sets. The average and maximum ambient temperatures in the hot year set are higher than in any other set. The average is 2°C higher for the hot-year set than for the training set, thus reflecting a distribution shift.

\autoref{tab: hot year} shows the metrics of the ITO and DFL models over the hot year data set. The results show a slight performance degradation for both models compared to the randomly sampled test set. Performance degradation is expected because the hot year data distribution is purposely different from the previous distribution.\newline
The hierarchical loss stands at 685 (+5.1\% compared to the test set) and 271 (+7.1\%) for the ITO and DFL models, respectively. The error standard deviation remains virtually unchanged. The mean, however, worsens for the ITO model but improves for the DFL model. This results in a greater cost error for the ITO but not for the DFL.\newline
Overall, DFL outperforms ITO and is less affected by input distribution shift than the ITO model.

\begin{table}[ht]
\begin{tabular}{@{}lll@{}}
\toprule
                                     & ITO   & DFL  \\ \midrule
Hierarchical loss                    & 685   & 271  \\
MAE (kW)                             & 2.76  & 2.95 \\
MSE (kW\textsuperscript{2})          & 18.3  & 15.8 \\
Error mean (kW)                      & -1.96 & 0.1  \\
Error std (kW)                       & 2.39  & 2.39 \\
Expected cost (\euro) & 79    & 467  \\
Ex-post cost (\euro)  & 500   & 482  \\
Cost error (\euro)    & 420   & 15   \\ \bottomrule
\end{tabular}
\caption{Metrics of the Identify-Then-Optimize (ITO) model and the Decision-Focused Learning (DFL) model over a dataset representing a distribution shift towards higher ambient temperature (\textrm{i.e.}, hot year).}
\label{tab: hot year}
\end{table}

\section{Discussion}
\label{sec: discussion}
The proposed \acrshort{dfl} strategy outperforms the \acrshort{ito} approach. The naive \acrshort{ito} framework that consists in \acrshort{mse} minimization over historical data leads to extremely poor dynamic model. In particular, the \acrshort{rc} model is unable to appropriately fit the operating areas relevant for the control policy. This leads to potentially severe electric power scheduling underestimation resulting in unexpectedly high ex-post bills. In the case study, the ex-post cost was on average 6 times higher than the expected cost for the \acrshort{ito} model. This raises important questions about the use of \acrshort{rc} model for the day-ahead \acrshort{hvac} scheduling, at least within an \acrshort{ito} framework. In contrast, the proposed DFL approach significantly improves both the accuracy of the bill prediction and the cost reduction. All error metrics show improvement, except for the \acrshort{mae}. However, the \acrshort{dfl} approach is much more computationally demanding. The \acrshort{dfl} duration before triggering the early stopping is about 7 hours compared to 16 minutes for the \acrshort{ito} approach. This is due to the \acrshort{dfl} need to solve, for each sample, the day-ahead \acrshort{hvac} scheduling optimization problem and simulating the results. Furthermore, software for \acrshort{dfl} are not as mature as the ones for supervized learning. Future software development for \acrshort{dfl} will likely harvest the power of parallel computing and new hardware leading to significant time reduction. In this specific case, due to the linear formulation of the optimization, the computational burden mainly lies with the simulation. We recommend considering using a reduced-order physics model of the simulator in the initial training steps.\newline
Finally, we showed that the model is robust to an input data shift towards higher temperature.

\section{Conclusion}
\label{sec: conclusion}

This paper presents a new framework that enables simultaneous identification and control (or planning) of a complex system. This is of particular interest for black-box systems in which dynamics change radically once the control policy is applied, thus making the historical observation little informative. The method exploits recent advances in decision-focused learning, and more specifically, in the calculation of the gradient of cone programs. The gradient of the convex optimization control policy output (\textrm{i.e.}, the command) with respect to the policy parameters (e.g., the system dynamics model parameters) can be computed. Because a key constraint of the control policy is being learned, the feasibility domain evolves at each gradient descent step. We handle the potential infeasibility by relaxing the constraints on the system states and pre-learning the uncertain parameters on historical data. Furthermore, we propose a new type of loss function that bypasses the need for the system to be differentiable. Not only is this necessary for black-box systems, but also for actual physical systems that cannot be easily modeled.
\par
We apply the proposed approach to the day-ahead \acrshort{hvac} scheduling of a 15-zone office located in Denver, CO. This case study showcases the efficient learning provided by our framework, and the ability to design complex task-aware loss functions to reflect the impact of the parameter misestimation on the objective value. The results show unambiguously the added value of simultaneous system identification and planning by displaying a lower ex-post cost along with a more accurate price forecast and a reduced error on the day-ahead  \acrshort{hvac} scheduling.
\par
Future work might consider building a surrogate model to bypass the simulation in the early \acrshort{dfl} steps. More detailed models for building dynamics should also be investigated. Such models will likely turn the optimization formulation into a mixed integer linear program that will bring new challenges in the computation of the gradient. In addition, the appropriate number of representative scenarios necessary for \acrshort{dfl} training should be explored. A trade-off must be found between a good generalization of the \acrshort{dfl} model to the whole initial dataset and the substantial computational time that can be saved. Lastly, future work should quantify the impact of a change in building use that results in new internal heat load, occupancy pattern, or temperature setpoints, and thus, HVAC operating points. 

 \section{Acknowledgements}
 Pietro Favaro is an FNRS-F.R.S. Research Fellow (grant number FC 49537) and Research Fellow of the Belgian American Educational Foundation (B.A.E.F.).

\bibliographystyle{ACM-Reference-Format}
\renewcommand{\UrlBreaks}{\do\/\do-}
\bibliography{ref}

\end{document}